\documentclass[preprint,prb,showkeys]{revtex4}
\usepackage{graphicx}
\usepackage{amsmath,array,graphicx}

\usepackage{color}
\usepackage{multirow}
\begin{document}
\title{Large Enhancement and Tunable Band Gap in Silicene by Small Organic Molecule Adsorption}
\author{Thaneshwor P. Kaloni}
\email{thaneshwor.kaloni@umanitoba.ca, +1-204-592-2900}
\author{Georg Schreckenbach}
\author{Michael S. Freund}
\affiliation{Department of Chemistry, University of Manitoba, Winnipeg, MB, R3T 2N2, Canada}

\begin{abstract}

Adsorption of eight organic molecules (acetone, acetonitrile, ammonia, benzene, methane, methanol, ethanol, and toluene) onto silicene has been investigated using van der Waals density functional theory calculations (DFT-D). The calculated values of the adsorption energies vary from $-0.11$ eV to $-0.95$ eV. Quantitatively, these values are higher than the corresponding adsorption energies of the molecules adsorbed on graphene. In addition, electronic structure calculations have been performed. The obtained values of the band gap range from 0.006 eV to 0.35 eV for acetonitrile to acetone, respectively. Furthermore, the effective mass of the electron is estimated and found to be comparatively small, which is expected to result in high electron mobility. In addition, we study the effect of Li atoms doped in pristine and acetone adsorbed silicene. In particular, we focus on the variation of the adsorption energy with respect to the number of Li atoms in the systems. Our results suggest new approaches for the use of silicene molecular-based energy storage and conversion as well as electronic devices. 

\end{abstract}

\maketitle

\section{Introduction}

Silicene is the subject of considerable attention due to its exotic electronic properties and promising applications in Si nanoelectronics that are similar to those of graphene.\cite{feng,antoine,jose} The similarity between silicene and graphene is due to the fact that Si and C belong to the same group IV in the periodic table. Generally, $sp^3$ hybridization is favourable for Si whereas $sp^2$ hybridization is energetically favourable in case of graphene. Experimentally, the growth of silicene on Ag(110) \cite{padova,padova1}, Ag(111) \cite{vogt}, ZrB$_2$ \cite{antoine}, and Ir(111) \cite{meng} have been reported. In addition, silicene on semiconducting substrates ($h$-BN and SiC) \cite{kaloni-scirep,kaloni-prb,liu3} have been studied and demonstrating that these substrates could be promising for producing free-standing silicene. Based on first-principles calculations, it has been shown that the electronic structure of silicene is particularly interesting due to the formation of graphene-like dispersion at the Fermi level (Dirac-like cone).\cite{olle} The vibrational properties measured from phonon dispersion calculations show that the silicene lattice can be stable up to a biaxial strain of 17\%.\cite{kaloni-jap} In addition, the effect of the strain on the mechanical properties of the free-standing silicene has been demonstrated to have larger intrinsic stiffness as compared to unstrained case. \cite{Pei} The structural and magnetic properties of 3$d$ transition metal doped silicene have been reported and it has been found that transition metals are strongly bound to the silicene sheet with higher binding energies compared to graphene.\cite{PRB,peeters} Moreover, adsorption of Fe and Cr on planar and buckled silicene has been studied and it was found that Fe and Cr bind strongly to buckled silicene as compared to the planar one.\cite{Bui} Based on the experimental and theoretical studies, it is expected that silicene can easily be integrated in electronic devices.  

The interactions between graphene and different chemical functional groups have potential applications in graphene-based devices.\cite{Krepel,nat.mater.} The physisorption of small molecules on graphene is of considerable interest since absorbents enhance the chemical reactivity of graphene.\cite{Chen,Wang} It has also been demonstrated that several organic molecules have potential to modify the electronic properties of graphene-based systems \cite{jacs1}, and that introducing dopants can enhance the interaction, which could be relevant for sensing devices.\cite{Chen,kaloni-carbon} Dopants in silicene also significantly modify the electronic and magnetic behaviour of the pristine silicene.\cite{kaloni-prb1,RRL} In particular, transition metal doped silicene can host a quantum anomalous Hall effect, which is highly desirable for spintronic applications.\cite{kaloni-prb1}

A large enhancement of spin-orbit coupling of approximately 2.5 meV has been observed for hydrogen molecule adsorbed graphene, which is sufficient to realize the spin Hall effect.\cite{nature} Recently, the molecular adsorption and its effects on graphene have been studied in both experiment and theory.\cite{jacs} Inverse gas chromatographyhas been used to investigate the adsorption energies and enthalpies of molecules onto graphene. Excellent agreement between experiment and computational approach has been demonstrated. This kind of study is key for understanding graphene's potential for energy storage applications as well as molecular-based graphene devices. To the best of our knowledge, studies of the adsorption of molecular species onto silicene have not been reported. In this paper, we address the issue of the adsorption of molecules onto silicene. Specifically, we investigate acetone, acetonitrile, ammonia, benzene, methane, methanol, ethanol, and toluene adsorbed silicene. In particular, the structural and electronic properties are addressed to quantify the adsorption energy and the effect on the electronic structure. In addition the effective mass of the electron has been estimated and indicates larger mobilities in these systems. Furthermore, the effect of doping Li atoms in pristine and acetone adsorbed silicene is addressed. In particular, the variation of the adsorption energy with the number of Li atoms has been focused. Our results show that the systems under consideration are promising for tuning the band gap and hence could have great potential for silicene based devices.  

\section{Computational details}
Unless otherwise noted, all the calculations were performed using density functional theory (DFT) within the generalized gradient approximation in the Perdew, Burke, and Ernzerhof parametrization \cite{pbe} as implemented in the PWSCF code.\cite{paolo} We include the van der Waals interaction (DFT-D) in order to achieve an accurate description of the dispersion.\cite{grime} A high plane wave cutoff energy of 816 eV and a Monkhorst-Pack $24\times24\times1$ k-mesh were used for all the structures under consideration. A $4\times4\times1$ supercell of silicene with a lattice constant of $a=15.44$ \AA\ was employed. In addition, a vacuum layer of 20 \AA\ was used in order to avoid artificial interactions due to the periodic boundary conditions. The supercell contains 32 Si atoms and a single molecule (acetone, acetonitrile, ammonia, benzene, methane, methanol, ethanol, and toluene) adsorbed on it. Based on the similar kinds of the modelling in graphene, this supercell is sufficient for our calculations.\cite{jacs} The adsorption density is low enough such that interaction between the molecule and its periodic images is prevented. Atomic positions were optimized until all forces had converged to less than 0.001 eV/\AA. 

\section{Results and Discussion}
\subsection{Structural Description}
\begin{figure}[h]
\includegraphics[width=0.5\textwidth,clip]{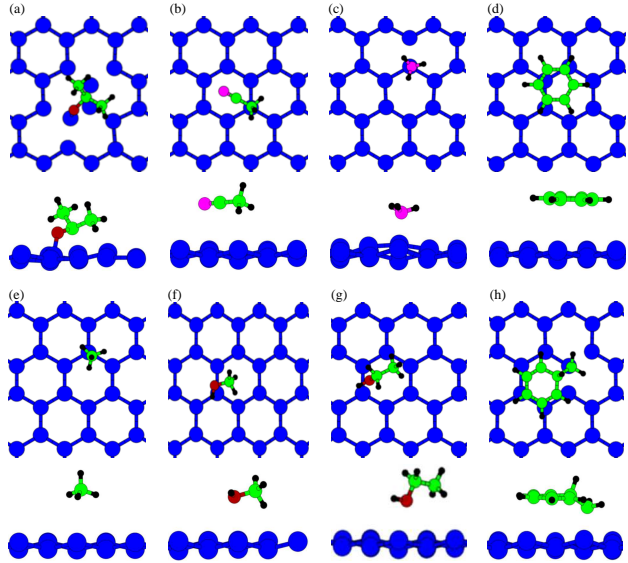}
\caption{Top and side views of the optimized structures for (a) acetone, (b) acetonitrile, (c) ammonia, (d) benzene, (e) methane, (f) methanol, (g) ethanol, and (h) toluene, where blue, green, black, violet, and red spheres represent Si, C, H, N, and O atoms, respectively.}

\end{figure}

\begin{table}[ht]
\caption{Adsorption energy (in eV) calculated by employing GGA-PBE (at Si-site) and DFT-D (Si-site, hollow site, and Si$-$Si bonds).}
\begin{tabular}{|c|c|c|c|c|}
\hline
system& GGA-PBE&\multicolumn{3}{|c|}{\multirow{1}{*}{DFT-D}}\\
\hline
       & &Si-site&H-site &Si$-$Si-site \\
 \hline
acetone& $-0.523$& $-0.614$&$-0.583$&$-0.529$      \\
\hline
acetonitrile& $-0.132$&$-0.171$&$-0.149$&$-0.113$   \\
\hline
 ammonia& $-0.293$& $-0.412$&$-0.382$&$-0.335$       \\
\hline
 benzene& $-0.219$& $-0.241$&$-0.296$&$-0.219$       \\
\hline
methane& $-0.056$& $-0.083$&$-0.061$&$-0.058$        \\
\hline
methanol& $-0.092$& $-0.111$&$-0.091$&$-0.085$       \\
\hline
ethanol& $-0.147$& $-0.201$&$-0.163$&$-0.136$        \\
\hline
toluene& $-0.764$& $-0.849$&$-0.949$&$-0.812$        \\
\hline
\end{tabular}
\label{table:1}
\end{table} 

The structures under consideration are illustrated in Fig.\ 1 (a-h). Note that three possibilities adsorption sites are considered for the calculations, which are categorized as the molecule being on top of a Si atom, on top of Si$-$Si bonds, and on the hollow site. The energetically most favourable cases are studied, which are listed in Fig.\ 1 with the corresponding energies presented in Table 1. In addition, we have checked the other possibilities by rotating the molecule and find that the rotated structures are energetically less favourable. It is found that most of the molecules prefer to attach on top of Si atoms, while benzene and toluene prefer to attach on the hollow site, see Table 1. The averaged perpendicular distance between the Si atoms called buckling ($\Delta$), the hexagonal angle between Si to the next Si to the next atom ($\theta$), and Si$-$Si bond lengths are hugely modify after adsorption of the molecules in silicene. The $\Delta$ varies from 0.39-0.53 \AA\ to 0.42-0.48 \AA\ for acetone to toluene, respectively, as summarized in Table 2. Similarly, the angle and Si$-$Si bond lengths as a result of adsorption of the molecules and are found to be 112-119$^{\circ}$ to 113-117$^{\circ}$ and 2.26-2.36 \AA\ to 2.26-2.29 \AA\ for acetone to toluene, see Table 2. These values are in good agreement with transition metal doped silicene with and without substrate.\cite{kaloni-prb,kaloni-prb1,Yang} In addition, after the adsorption of the molecules, the C$-$C bond length are 1.52 \AA, 1.45 \AA, 1.39 \AA, 1.51 \AA, and 1.37-1.51 \AA\ for acetone, acetonitrile, benzene, ethanol, and toluene, respectively. In case of acetone a O$-$Si bond length of 1.69 \AA\ was obtained, which indicates a stronger binding of acetone with the host silicene sheet with significant structural reconstruction close to the adsorbed region and a O$-$C bond length of 1.45 \AA. For acetonitrile a N$-$C bond length of 1.16 \AA\ is obtained. In the case of ammonia a N$-$H bond length of of 1.03 \AA\ is found. For methanol a bond length of C$-$O=1.43 \AA\ and O$-$H=0.97 \AA\ and in ethanol a bond length of O$-$C=1.44 \AA\ and O$-$H=0.97 \AA\ are realized. The distance ($d$) between the silicene sheet and molecules varies from 3.04 \AA\ to 3.41 \AA, similar to experimental and theoretical reports for graphene.\cite{jacs} The fact is that the modification in the structural parameters and local distortion greatly influence the electronic properties as well as the chemical reactivity of the silicene.\cite{Chong,Ongun,afm}

\begin{table}[htb]
\centering\setlength{\tabcolsep}{.3\tabcolsep}
\caption{Pristine silicene and adsorbed molecule, band gap, structural parameters, the effective mass of the electron (in $m_e$), and the amount of the charge transfer from the adsorbent (in electron) for the systems under study.}  
\begin{tabular}{|c|c|c|c|c|c|c|c|c|}
\hline
system&E$_{gap}$ (eV)&$d$ (\AA) &$\theta$ $^{\circ}$&Si$-$Si (\AA) & $\Delta$ (\AA)&C$-$C (\AA)&$m_e^*$&$Q$\\
\hline
silicene&0.002         &--        &116                 &2.28         &0.46 &--&0.001&--\\
\hline
acetone&0.351 &3.04 &112-119 & 2.26-2.36&0.39-0.53 &1.52  &0.166&0.483\\
\hline
acetonitrile& 0.006&3.38 &115-116 &2.27-2.28 &0.44-0.52 &1.45 &0.003&0.008     \\
\hline
 ammonia& 0.103&3.42 &113-117 &2.27-2.32 &0.45-0.51 &--  &0.049&0.167 \\
\hline
 benzene& 0.007&3.35 &114-116 &2.27-2.28 &0.42-0.50 &1.39 &0.003&0.009     \\
\hline
methane&0.002 &3.10 &113-117 &2.27-2.28 &0.42-0.48&--   &0.001&0.004   \\
\hline
methanol&0.011 &3.48 &115-117 &2.27-2.28 &0.45-0.50 &--   &0.005&0.012  \\
\hline
ethanol&0.017 &3.22 &114-116 &2.27-2.29 & 0.44-0.49&1.51 &0.008&0.019    \\
\hline
toluene&0.208 &3.41 &113-117 &2.26-2.29 & 0.42-0.48&1.37-1.51 &0.098&0.291    \\
\hline
\end{tabular}
\label{table:2}
\end{table}

\subsection{Adsorption energy}
Our structural analysis shows that in some cases adsorption results in major, but in some cases only in minor changes in the shape of the adsorbed molecule as well as that of the host silicene sheet. This can have strong effects on the adsorption energy. The adsorption energy of the molecules on silicene is calculated as $E_{ad} = (E_{molecule} + E_{silicene}) - E_{total}$, where $E_{molecule}$ is the total energy of the isolated molecule, $E_{silicene}$ is the total energy of a $4\times4\times1$ pristine silicene supercell, and $E_{total}$ is the total energy of the molecule adsorbed silicene. The obtained values of $E_{ad}$ for the different molecules on silicene range from $-0.083$ eV to $-0.949$ eV, as summarized in Table 1. The calculated data for the $E_{ad}$ shows that methane is weakly adsorbed, while toluene is far more strongly adsorbed than all other systems under study. Theoretically, the adsorption energy of ammonia on graphene has been reported to be 0.11 eV \cite{peng1}; this value is approximately four times lower than ammonia adsorbed on silicene. This indicates that silicene is an extremely good candidate compared to graphene to detect or sensor harmful gases, like ammonia. In general our data for the adsorption energy shows that these molecules are more strongly bound to silicene as compared to graphene and hence silicene could be the better host material for these molecules, which makes it a potential candidate material in sensor as well as electronic device applications. 

\subsection{Electronic structure}
\begin{figure}[h]
\includegraphics[width=0.5\textwidth,clip]{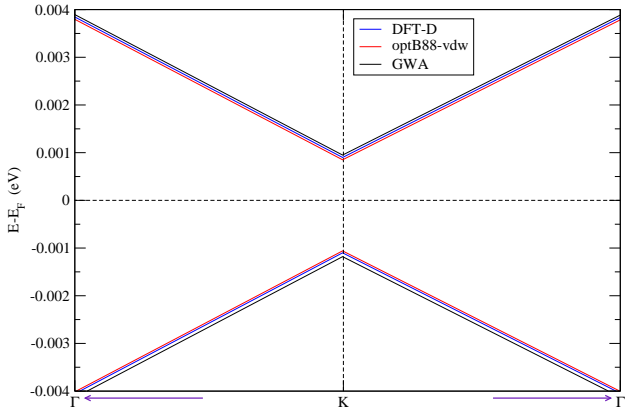}
\caption{Electronic band structure of pristine silicene using DFT-D, optB88-vdW functional, and the GW approximation. The arrows point out the direction of the $\Gamma$ point. }
\end{figure}

It is well known that silicene is a narrow gap semiconductor, where the $p_z$ and $p_z^*$ orbitals give rise to $\pi$ and $\pi^*$ ands forming Dirac cones at the K and K$'$ points. In order to check the influence of the different functionals on the electronic structure, in particular the band gap at the K point, the electronic structure of pristine silicene has been calculated with three different approaches; (i) DFT-D, (ii) optB88-vdW functional, and (iii) the GW approximation using the VASP package \cite{vasp}, see Fig.\ 2. Here ``opt" is the the optimised exchange functionals \cite{opt}, G refers to the one-electron Green’s function, and W refers to the screened Coulomb interaction, and their product ``GW" is the quasi-particle self-energy corrections methods.\cite{gw} It is found that the band gap is 1.92 meV, 1.95 meV, and 1.99 meV by using DFT-D, the optB88-vdW functional, and the GW approximation, respectively. The obtained result agrees well with the available reports \cite{Seixas,Vargiamidis2} and hence in the following only the DFT-D calculations have been taken into account. Modifications in the electronic structure can be achieved by applying strain/electric fields \cite{kaloni-jap,kaloni-scirep}, and doping.\cite{kaloni-prb1} However, the impact of molecular adsorption on the electronic structure has not been explored. The electronic structure of the systems under consideration are illustrated in Fig.\ 3. The electronic structure of acetone adsorbed silicene is strongly modified resulting in E$_{gap}=0.351$ eV, see Fig.\ 3(a). This gap opening can be attributed to the strong structural distortion, which breaks the structural symmetry. Such a modification has also been found in case of the transition metal decorated silicene.\cite{kaloni-prb,kaloni-prb1} Due to the significant band gap as compared to pristine silicene, the system is expected to be of potential use for electronic device applications such as field effect transistors. The band gap can be further increased by the application of an electric field such that the system could be promising for opto-electronic devices. Due to the weak interaction between silicene and acetonitrile, a small band gap of 0.006 eV is achieved, see Fig.\ 3(b), which is still about 4 times higher than that of pristine silicene ($\sim1.9$ meV). In case of ammonia adsorbed silicene, a band gap of 0.103 eV is obtained, see Fig.\ 3(c). The fact is that silicene is locally distorted around the ammonia adsorbed areas. The band structure of benzene adsorbed silicene is addressed in Fig.\ 3(d). Due to a weak interaction between silicene and benzene a band gap of 0.007 eV is obtained. A band gap of 0.002 eV is obtained in case of methane adsorbed silicene, see Fig.\ 3(e), which is similar to that of pristine silicene. This indicates only a minimal interaction between silicene and methane. In addition, larger band gaps of 0.011 eV, 0.017 eV, and 0.208 eV are obtained for methanol, ethanol, and toluene adsorbed silicene, see Figs.\ 3(f), 3(g), and 3(h), respectively.

\begin{figure}[h]
\includegraphics[width=0.5\textwidth,clip]{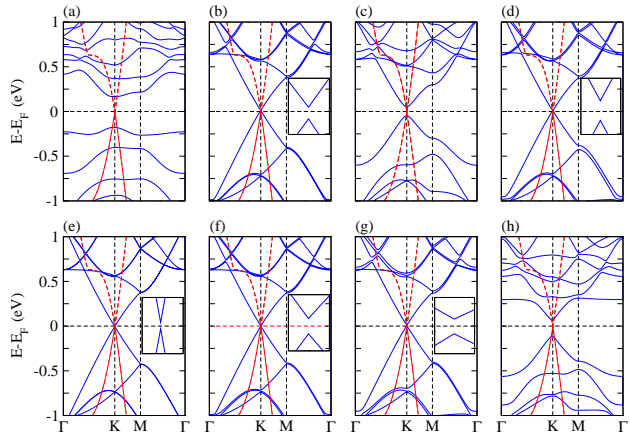}
\caption{Electronic band structure for (a) acetone, (b) acetonitrile, (c) ammonia, (d) benzene, (e) methane, (f) methanol, (g) ethanol, and (h) toluene adsorbed silicene. The red dashed lines represent the band structure for pristine silicene and the inset shows the zoomed areas at the K point around the Fermi level.} 
\end{figure}

A clear trend is observed between the band gap opening and the adsorption energy. While the correlation between band gap and adsorption energy is not perfect, we note that stronger adsorption provides a larger band gap opening and weaker adsorption provides a smaller band gap opening, which can be clearly seen from the data presented in Fig.\ 4(A), this behaviour agrees well with the band gap engineered by adsorption of aromatic molecules on graphene.\cite{chang1} For instance, the adsorption energy of acetone is highest and acetone-adsorbed silicene has a very large band gap. The band gap opening can be understand by looking at the charge transfer from the adsorbent to the host silicene. Charge transfer between adsorbent molecule and silicene provides an 
internal electric field, which indeed breaks the sublattice symmetry in silicene and opens a band gap at the Dirac point (at the K-high symmetric point). The internal electric field can be understood by using a parallel plate capacitor model, which has been already well established for 2D systems, such as graphene \cite{Ashwin,kaloni-jmc1}, silicene \cite{kaloni-scirep,scirep2,Ni-nlett}, and germanene.\cite{physicae} This can be calculated by using $E_{field}=\frac{2Q} {\epsilon_0 a^2 sin(\pi/3)}$, where $Q$, $\epsilon_0$ and $a$ are the L\"owdin transferred charge, vacuum permittivity, and silicene lattice constant, respectively. Previous theoretical calculations already demonstrated that the band gap of silicene/germanene increases linearly with increasing external electric field \cite{Ni-nlett}. In our calculations we find that the largest amount of the charge transfer (0.483 electron) from the adsorbent to silicene provides the highest value of the band gap (for acetone adsorption), and the smallest amount of charge transfer (0.004 electron) for methane induces a minimal value of the band gap, see Fig.\ 4(B). The calculated values of the charge transfer are summarized in Table 2. Moreover, the obtained value of the band gap can be enhanced by applying an electric field, which is relevant for the design of molecular based electronic devices. Experimentally, molecular based electronic devices, in particular field effect transistors, have already been designed for molecular oxygen adsorbed graphene \cite{he,Standley} and fluorinated graphene \cite{Robinson} and hence it is expected that molecule adsorbed silicene would also be useful to construct silicene based devices such as field effect transistors. 

\begin{figure}[h]
\includegraphics[width=0.5\textwidth,clip]{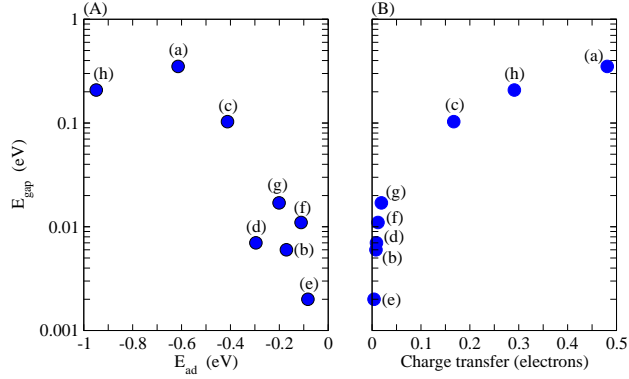}
\caption{(A) The adsorption energy versus the band gap and (B) the charge transfer versus the band gap (logarithmic scale) for (a) acetone, (b) acetonitrile, (c) ammonia, (d) benzene, (e) methane, (f) methanol, (g) ethanol, and (h) toluene adsorbed silicene.} 
\end{figure}
\subsection{Quantitative analysis: Effective mass and mobility}
The energy dispersion relation for the gapped 2D systems, where charge carriers behave as massive Dirac fermions, graphene and silicene for example, can be obtained \cite{Jiang} by $ E(k)=\pm v_F\sqrt{k_x^2 + k_y^2 + m_e^*2v_F^2}$. The $\pm$ represents the conduction and valence bands, $k_x$ and $k_y$ are wave vectors along the x- and y-directions, and $m_e^*$ is an effective mass of the electrons. It is calculated as $m_e^* = E_{gap}/2v_F^2$, where $E_{gap}$ is the band gap and $v_F$ is the Fermi velocity, which is $1.3\times10^6$ ms$^{-1}$ for pristine silicene.\cite{Meng} It has been demonstrated that the layer-layer interaction is the key factor in order to control/tune the band gap in graphene systems, however, the value of the $m_e^*$ is enhanced largely as a function of the increment in the band gap, which has poor performance for device application because of smaller mobility. Therefore, it is necessary to keep the value of $m_e^*$ as small as possible.\cite{udo-apl} The calculated values of the effective mass of the electron are found to be 0.001 $m_e$ to 0.166 $m_e$, see Table 2. These values are significantly smaller as compared to doped graphene \cite{udo-apl} and double-gate graphene nanoribbon field-effect transistors.\cite{Kliros} Hence, doped silicene would provide larger mobility, which is good for the device performance. Interestingly, it is observed that the smaller effective mass for the molecule having smaller adsorption energy and vice-versa. The electron mobility can be evaluated as $\mu = e\tau/m_e^*$, where $\tau$ is the momentum relaxation time. At this point, the exact value of $\tau$ is unknown and hence it is not possible to estimate the exact value of $\mu$. However, the smaller values of the effective mass of the electron ensure that the $\mu$ in the doped silicene systems should be very high, even higher than those in the case of graphene and its derivatives. 

\subsection{Effect of Li doping on pristine versus acetone adsorbed silicene}
\begin{figure}[h]
\includegraphics[width=0.5\textwidth,clip]{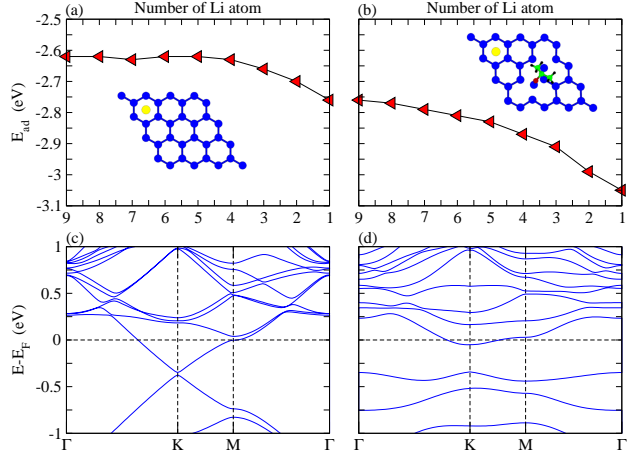}
\caption{Prototypical structures and variation of the adsorption energy with respect to number of Li atom in silicene in case of (a) pristine and (b) acetone adsorption. The corresponding electronic band structures are presented in (c) and (d), respectively.} 
\end{figure}

In this section, we explore the effect of Li atom doping in pristine silicene and in acetone adsorbed silicene. We start from single Li atom doping in one of the hollow sites, see the prototype structures in Figs.\ 4(a) and (b). We chose acetone adsorbed silicene because the adsorption energy is highest in this case out of all the eight molecules discussed before and we aim to understand how the adsorption energy and electronic property influenced by the doping of Li atoms. The Li atom occupies the hollow site without creating any distortion to the Si hexagon (no modification in the Si$-$Si bond lengths). The average distance of the Li atom to the silicene sheet is about 1.72 \AA\ and 1.68 \AA\ in case of pristine and acetone adsorbed silicene, respectively. The obtained distance agrees well with the experimentally and theoretically observed value for Li doped graphene \cite{Virojanadara,kaloni-cpl2,kaloni-epl2} and theoretical prediction for Li doped silicene.\cite{scirep2} The obtained value of the adsorption energy for a single Li atom doped silicene is found to be $-$2.76 eV, this value agrees well with Li doping in nitrogen mediated silicene.\cite{scirep2} Furthermore, the adsorption energy decreases monotonically with increasing the number of Li atoms in silicene, see Fig.\ 5(a), which agrees well with O doped graphene \cite{udo-jmc}. It has been predicted that Li doped monolayer/bilayer graphene supported by the SiC substrate can have high potential to construct nano-scale Li-ion batteries.\cite{Virojanadara} Similarly, Li doped silicene can also be a potential candidate for Li-ion batteries.\cite{Virojanadara} The lack of structural distortion discussed above is a significant factor here also, since structural reconstruction with Li motion during charging/discharging is undesirable. The adsorption energy of single Li atom doping in acetone adsorbed silicene is found to be $-$3.05 eV, higher than for Li doped pristine silicene. 

This indicates that the acetone molecule has a significance influence on the adsorption of Li atoms, which also indicates that one can use molecular adsorption (acetone for example) in order to achieve higher energy for Li adsorption. In addition, the strength of adsorption of an acetone molecule in the presence of a Li atom is about 5 times higher than that of adsorption of acetone on pristine silicene, which means that depending on the need, one can include Li atom during the acetone molecule adsorption on silicene. The value of the adsorption energy decreases with increasing the number of Li atoms, see Fig.\ 5(b). The reduction of the adsorption energy can be understood by possible Li-Li interactions. In Fig.\ 5(c), the electronic structure of Li doped pristine silicene along the $\Gamma$-K-M-$\Gamma$ path is shown. The Dirac-like cone in pristine silicene shifts below the Fermi level by 0.36 eV by providing a gap of 0.03 eV. The shifting of the Dirac-like cone can be attributed to the charge transfer from the Li atom to the Si atoms of the silicene hexagon, amounting to 0.83 electrons. In addition, the opening of the band gap is due to the hexagonal symmetry breaking.\cite{kaloni-epl2,scirep2} Moreover, a gap of 0.27 eV below the Fermi level by 0.06 eV is obtained in case of Li doping on acetone molecule adsorbed silicene (see Fig.\ 5 (d)), which is in fact due to the same reason as for Li doping on pristine silicene. Thus, this material can also have potential for use in molecule based Li-ion batteries.

\section{Conclusion}
In conclusion, using van der Waals density functional theory (DFT-D), the structural and electronic properties of acetone, acetonitrile, ammonia, benzene, methane, methanol, ethanol, and toluene adsorbed silicene have been studied. The obtained values of the adsorption energy vary from $-0.11$ eV to $-0.95$ eV for acetone to toluene, respectively. These values are higher than those for similar molecules adsorbed graphene, which indicates that silicene could be a better candidate as compared to graphene to detect the organic molecules and hence to tune the electronic properties. In addition, band structure calculations were performed. The calculated band gaps range from 0.006 eV to 0.35 eV for acetonitrile to acetone, respectively. Furthermore, the gap can be enhanced by applying an electric field, which, indeed, could be potential to design the molecular based electronic devices. Our calculated values of the effective mass of the electron are found to be smaller as compared to the graphene counterpart and hence we expect larger mobility, which should be of great importance for silicene-based electronic devices. In addition, the effect of Li atoms in pristine and acetone adsorbed silicene is discussed. It is found that the adsorption energy is reduced with increasing the number of the Li atoms in the systems. The results show that Li doped pristine/acetone adsorbed silicene can be an alternative candidate for Li-ion batteries as Li doped graphene has a great potential for constructing Li-ion batteries.

\section{Acknowledgement}
GS acknowledges funding from the Natural Sciences and Engineering Council of Canada (NSERC, Discovery Grant). MSF acknowledges support by the Natural Sciences and Engineering Research Council (NSERC) of Canada, the Canada Research Chair program, Canada Foundation for Innovation (CFI), the Manitoba Research and Innovation Fund, and the University of Manitoba.

\end{document}